\documentclass[aps,prd,twocolumn,superscriptaddress]{revtex4}
\usepackage{epsfig,epsf}
\usepackage{amsmath}
\usepackage{amsthm}
\usepackage{amsfonts}
\usepackage{amssymb}
\usepackage{dsfont}
\usepackage{multirow}
\usepackage{appendix}
\usepackage{slashed}
\usepackage[active]{srcltx}
\usepackage{psfrag}

\begin{document}

\title{Probing an axial-vector tetraquark $Z_s$ via its semileptonic decay $%
Z_s \to X(4274)\overline{l} \nu_l$}
\date{\today}
\author{H.~Sundu}
\affiliation{Department of Physics, Kocaeli University, 41380 Izmit, Turkey}
\author{B.~Barsbay}
\affiliation{Department of Physics, Kocaeli University, 41380 Izmit, Turkey}
\author{S.~S.~Agaev}
\affiliation{Institute for Physical Problems, Baku State University, Az--1148 Baku,
Azerbaijan}
\author{K.~Azizi}
\affiliation{Department of Physics, Do\v{g}u\c{s} University, Acibadem-Kadik\"{o}y, 34722
Istanbul, Turkey}
\affiliation{School of Physics, Institute for Research in Fundamental Sciences (IPM),
P.~O.~Box 19395-5531, Tehran, Iran}

\begin{abstract}
The semileptonic decays of the open charm-bottom axial-vector tetraquark $%
Z_{s}=[cs][\overline{b}\overline{s}]$ to $X(4274)\overline{l} \nu_l$, $l=e,
\mu, \tau $ are explored by means of the QCD three-point sum rule method. 
Both $Z_{s}$ and $X(4274)=[cs][\overline{c}\overline{s}]$ are treated as
color sextet diquark-antidiquark states. The full width of the decays $Z_s
\to X(4274)\overline{l} \nu_l$ is found. Obtained predictions for $%
\Gamma(Z_s \to X(4274)\overline{l} \nu_l)$ demonstrate that, as in the case
of the conventional hadrons, the semileptonic transitions form a very small
part of its full width.
\end{abstract}

\maketitle


\section{Introduction}

\label{sec:Int} 
The hadronic inclusive and exclusive processes, their experimental
investigation and interpretation within existing theories and models are
sources of valuable information on structures and properties of elementary
particles. The increasing precision of experimental studies allows one not
only to measure parameters of the well known baryons and mesons, but also to
discover new multiquark or exotic states. These states were theoretically
predicted already in the context of the quark model \cite%
{Jaffe:1976ig,Weinstein:1982gc}, but the first strong evidence for their
existence appeared only in 2003, when the Belle Collaboration announced
about the observation of the four-quark state $X(3872)$ \cite{Choi:2003ue}.
The narrow charmonium-like state $X(3872)$ was later confirmed independently
by different collaborations such as D0, CDF and BaBar experiments \cite%
{Abazov:2004kp,Acosta:2003zx,Aubert:2004ns}. During the time passed from
this discovery due to throughout investigations of $B$ meson decays, $%
e^{+}e^{-}$ and $\overline{p}p$ annihilations, $pp$ collisions and other
processes by the Belle, BaBar, BESIII, LHCb, D0 collaborations wide
information is collected on the masses, decay widths and quantum numbers of
the exotic particles. Now the exotic states observed and studied
experimentally constitute a new and broad family of $XYZ$ particles.

Considerable efforts were made also to understand the features of the exotic
states and calculate their parameters within existing theoretical models or
to invent new approaches for solving  unusual problems emerged with their
discovery. All theoretical methods and computational schemes of high energy
physics starting from bag and quark models and ending by sum rules
calculations were activated to meet challenges of a new situation. The
details of the performed theoretical and experimental investigations,
information on achievements and existing problems can be found in the
reviews Refs.\ \cite%
{Jaffe:2004ph,Swanson:2006st,Klempt:2007cp,Godfrey:2008nc,
Faccini:2012pj,Esposito:2014rxa,Chen:2016qju,Chen:2016spr,Esposito:2016noz}
and in references therein.

The theoretical papers devoted to exotic states are concentrated mainly on
studies of their internal quark-gluon structure, spin, parity and C-parity,
on calculations of their spectroscopic parameters using numerous approaches.
Strong decay channels of the exotic particles also attract the interests of
physicists, but progress achieved in this branch of investigations is
considerably modest if compared to the one made in other fields. There are
articles in the literature where the hadronic decays of the four-quark
(tetraquark) states were analyzed by means of different methods and partial
widths some of these modes were found. Among these papers it is worth noting
Refs.\ \cite%
{Brito:2004tv,Dias:2013xfa,Agaev:2016dev,Agaev:2016ijz,Dias:2016dme,Sundu:2017xct}%
, where strong decays of the tetraquarks were studied on the basis of the
sum rules method. In the framework of alternative approaches the similar
hadronic decays of the tetraquarks, as well as their radiative and dilepton
decay modes were considered also in Refs.\ \cite%
{Dong:2013iqa,Dong:2013kta,Gutsche:2014zda,Esposito:2014hsa,Chen:2015igx,Gutsche:2017twh}%
.

Recently, information on the magnetic dipole and quadrupole moments of some
of the tetraquarks calculated by employing QCD light-cone sum rules approach
became available \cite{Agamaliev:2016wtt,Ozdem:2017jqh,Ozdem:2017exj}. There
is an evident necessity to extend the type of investigated processes with
tetraquarks to gain more detailed information on their structure and decay
properties that may be checked in future experiments. This is also important
to build a reliable framework for further theoretical analyses. In the
present work we pursue namely this goal: we are going to calculate the width
of the semileptonic decay $Z_{s}\rightarrow X(4274)\overline{l}\nu _{l}$
using the standard methods of QCD three-point sum rules. This will allow us
not only to check consistency of the applied method but also to get first
estimates for the rates of the tetraquark's semileptonic decays.

The axial-vector state $Z_{s}=[cs][\overline{b}\overline{s}]$ belongs to the
class of the open charm-bottom tetraquarks and has the symmetric or
sextet-type color structure \cite{Chen:2013aba}. The spectroscopic
parameters of the scalar and axial-vector open charm-bottom color sextet
tetraquarks, as well as partial width of their strong decays were computed
in Refs.\ \cite{Agaev:2016dsg,Agaev:2017uky}. These exotic states have not
been seen in experiments yet, and still have a status of interesting but
hypothetical particles. On the contrary, the group of four $X$ resonances
was recently studied by the LHCb Collaboration, which reported its results
of analysis of the exclusive decays $B^{+}\rightarrow J/\psi \phi K^{+}$,
and confirmed the existence of the resonances $X(4140)$ and $X(4274)$ in the
$J/\psi \phi $ invariant mass distribution \cite{Aaij:2016iza,Aaij:2016nsc}.
The LHCb also discovered the heavy resonances $X(4500)$ and $X(4700)$ in the
same $J/\psi \phi $ channel. The collaboration measured masses and decay
widths of these states, and determined their spin-parities, as well. It
turned out, that the quantum numbers of $X(4140)$ and $X(4274)$ are $%
J^{PC}=1^{++}$, whereas the $X(4500)$ and $X(4700)$ are the scalar particles
with $J^{PC}=0^{++}$. But apart from this standard analysis the LHCb
Collaboration on the basis of the collected experimental information ruled
out a treating of the $X(4140)$ as $0^{++}$ or $2^{++}$ $D_{s}^{\ast
+}D_{s}^{\ast -}$ molecular states. The LHCb also emphasized that molecular
bound-states or cusps can not account for the $X(4274)$ resonance. This
information considerably restricts the possible interpretation of the $X$
states. Thus, in Ref. \cite{Maiani:2016wlq} they were identified as the
members of $1S$ and $2S$ multiplets of color triplet $[cs]_{s=0,1}[\overline{%
c}\overline{s}]_{\overline{s}=0,1}$ tetraquarks. In accordance with this
scheme $X(4140)$ was identified with the $J^{PC}=1^{++}$ level of the $1S$\
ground-state multiplet. Then the resonance $X(4274)$ is presumably a linear
superposition of two states with $J^{PC}=0^{++}$ and $J^{PC}=2^{++}$. The
heavy resonances $X(4500)$ and $X(4700)$ were included into the $2S$
multiplet as its $J^{PC}=0^{++}$ members. But besides the color triplet
multiplets there may exist a multiplet of the color sextet tetraquarks \cite%
{Stancu:2006st}, which also contains a state with $J^{PC}=1^{++}$. In other
words, the multiplet of the color sextet tetraquarks doubles a number of the
states with the same spin-parity, and the resonance $X(4274)$ may be
identified with the $J^{PC}=1^{++}$ member of this multiplet.

In our previous paper \cite{Agaev:2017foq} we studied the axial-vector
resonances $X(4140)$ and $X(4274)$ using the diquark-antidiquark picture for
their internal organization, and color triplet and sextet type currents to
interpolate $X(4140)$ and $X(4274)$, respectively. We computed their
spectroscopic parameters and decay widths. In the present work we will use
the information about the resonance $X(4274)$\ obtained in Ref.\ \cite%
{Agaev:2017foq}.

This work is structured in the following manner: In Sec.\ \ref{sec:SumRules}
we derive the QCD three-point sum rules for the transition form factors $%
G_{i}(q^{2})\,i=1,2,3,4$ which are important ingredients of our
calculations. In the next section we derive the differential decay rate $%
d\Gamma /dq^{2}$ and perform numerical analysis of the derived expressions.
First, we evaluate the sum rules for $G_{i}(q^{2})$, fit them by the
functions $F_{i}(q^{2})$ and finally calculate the decay width $\Gamma
(Z_{s}\rightarrow X(4274)\overline{l}\nu _{l})$, $l=e,\ \mu $ and $\tau $
that are kinematically allowed semileptonic decay channels of the tetraquark
$Z_{s}$. The last section contains an analysis of the obtained results and
our brief concluding notes. The lengthy expression for the correlation and
some other functions are removed to the Appendix.


\section{Sum rules for the transition form factors $G_{i}(q^{2})$}

\label{sec:SumRules}
\begin{figure}[tbp]
\centerline{
\begin{picture}(210,200)(-0,-0)
\put(-15,5){\epsfxsize8.2cm\epsffile{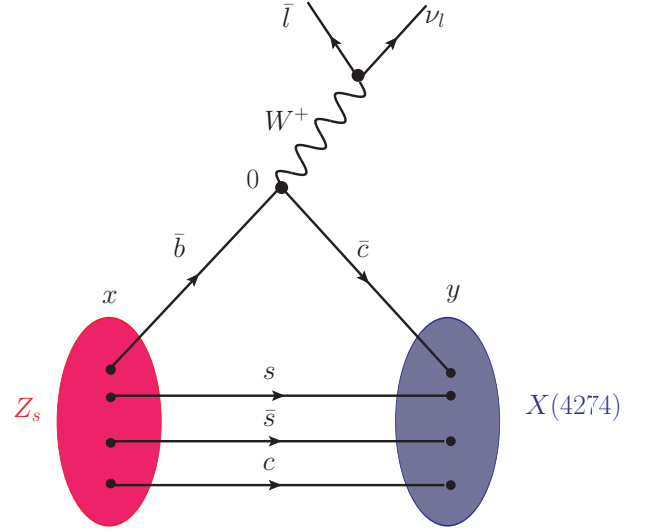}}
\end{picture}}
\caption{The diagram corresponding to the semileptonic decay $%
Z_{s}\rightarrow X(4274)\overline{l}\protect\nu _{l}$.}
\label{fig:Fgiagram}
\end{figure}

The semileptonic decay of the open charm-bottom tetraquark $Z_{s}$ to $%
X(4274)\overline{l}\nu _{l}$ proceeds through transition $\overline{b}%
\rightarrow W^{+}\overline{c}$ and decay $W^{+}\rightarrow \overline{l}\nu
_{l}$, as it is depicted in Fig.\ \ref{fig:Fgiagram}. The mass of the $Z_{s}$
state
\begin{equation}
m=7.30\pm 0.76\ \mathrm{GeV},  \label{eq:Mass1}
\end{equation}%
evaluated in Ref.\ \cite{Agaev:2017uky} is large enough, and it is evident
that all decays $l=e,\ \mu $ and $\tau $ are kinematically allowed processes.

The tree-level transition $\overline{b}\rightarrow \overline{c}$ can be
described using the effective Hamiltonian
\begin{equation}
\mathcal{H}^{\mathrm{eff}}=\frac{G_{F}}{\sqrt{2}}V_{bc}\overline{c}\gamma
_{\mu }(1-\gamma _{5})b\overline{l}\gamma ^{\mu }(1-\gamma _{5})\nu _{l},
\label{eq:EffH}
\end{equation}%
where $G_{F}$ is the Fermi coupling constant and $V_{bc}$ is the
corresponding element of the Cabibbo-Kobayashi-Maskawa (CKM) matrix. After
sandwiching the $\mathcal{H}^{\mathrm{eff}}$ between the initial and final
states we get the matrix element for the weak transition current%
\begin{equation}
J_{\alpha }^{\mathrm{tr}}=\overline{c}\gamma _{\alpha }(1-\gamma _{5})b,
\label{eq:TrCurr}
\end{equation}%
parameterized in terms of the form factors $G_{i}(q^{2})$

\begin{eqnarray}
&&\langle X(p^{\prime },\epsilon ^{\prime })|J_{\alpha }^{\mathrm{tr}%
}|Z_{s}(p,\epsilon )\rangle =\epsilon ^{\theta }\epsilon ^{\prime \beta }%
\left[ G_{1}(q^{2})g_{\theta \beta }(p+p^{\prime })_{\alpha }\right.  \notag
\\
&&\left. +G_{2}(q^{2})\left( q_{\theta }g_{\alpha \beta }-q_{\beta
}g_{\alpha \theta }\right) -\frac{G_{3}(q^{2})}{2m^{2}}q_{\theta }q_{\beta
}(p+p^{\prime })_{\alpha }\right]  \notag \\
&&+G_{4}(q^{2})\varepsilon _{\alpha \theta \rho \beta }\epsilon ^{\theta
}\epsilon ^{\prime \rho }(p+p^{\prime })^{\beta},  \label{eq:Vertex}
\end{eqnarray}%
where $m$ is the mass of the tetraquark $Z_{s},$ whereas by $(p,\epsilon )$
and $(p^{\prime },\epsilon ^{\prime })$ we denote the momenta and
polarization vectors of the $Z_{s}$ and $X(4274)$, respectively. In Eq.\ (%
\ref{eq:Vertex}) $q=p-p^{\prime }$ is the momentum transfer in the weak
transition process: $q^{2}$ changes within the limits $m_{l}^{2}\leq
q^{2}\leq (m-m_{X})^{2},$ where $m_{X}$ and $m_{l}$ are the masses of the
resonance $X(4274)$ and lepton $l$.

The transition form factors $G_{i}(q^{2})$ are key components in our
investigations. In order to derive the sum rules for these quantities we
begin from the calculation of the three-point correlation function
\begin{eqnarray}
&&\Pi _{\mu \alpha \nu }(p,p^{\prime })=i^{2}\int
d^{4}xd^{4}ye^{-ipx}e^{ip^{\prime }y}  \notag \\
&&\times \langle 0|\mathcal{T}\{J_{\nu }^{X}(y)J_{\alpha }^{\mathrm{tr}%
}(0)J_{\mu }^{^{\dagger }}(x)\}|0\rangle   \label{eq:CF}
\end{eqnarray}%
where $J_{\mu }(x)$ and $J_{\nu }^{X}(y)$ are the interpolating currents to
the $Z_{s}$ and $X(4274)$ $\ $states, respectively. They are given by the
following expressions:
\begin{equation}
J_{\mu }(x)=s_{a}^{T}C\gamma _{5}c_{b}\left( \overline{s}_{a}\gamma _{\mu }C%
\overline{b}_{b}^{T}+\overline{s}_{b}\gamma _{\mu }C\overline{b}%
_{a}^{T}\right) ,  \label{eq:Current1}
\end{equation}%
and%
\begin{eqnarray}
&&J_{\nu }^{X}(y)=s_{a}^{T}C\gamma _{5}c_{b}\left( \overline{s}_{a}\gamma
_{\nu }C\overline{c}_{b}^{T}+\overline{s}_{b}\gamma _{\nu }C\overline{c}%
_{a}^{T}\right)   \notag \\
&&+s_{a}^{T}C\gamma _{\nu }c_{b}\left( \overline{s}_{a}\gamma _{5}C\overline{%
c}_{b}^{T}+\overline{s}_{b}\gamma _{5}C\overline{c}_{a}^{T}\right) .
\label{eq:Curr2}
\end{eqnarray}%
In the equations above $C$ is the charge conjugation operator, $a$ and $b$
are the color indices.

The standard prescriptions of the sum rules require computation of the
correlation function $\Pi _{\mu \alpha \nu }(p,p^{\prime })$ employing both
the physical parameters of the involved particles, i. e. their masses and
couplings and also using the quark propagators, which give rise to $\Pi
_{\mu \alpha \nu }^{\mathrm{OPE}}(p,p^{\prime })$ in terms of quark, gluon
and mixed vacuum condensates. By matching the obtained results and invoking
the assumption on the quark-hadron duality it is possible to extract sum
rules and evaluate the physical parameters of interest.

Taking into account contribution arising only from the ground-state
particles one can easily write down $\Pi _{\mu \alpha \nu }^{\mathrm{Phys}%
}(p,p^{\prime })$ in the following form%
\begin{eqnarray}
&&\Pi _{\mu \alpha \nu }^{\mathrm{Phys}}(p,p^{\prime })=\frac{\langle
0|J_{\nu }^{X}|X(p^{\prime },\epsilon ^{\prime })\rangle \langle X(p^{\prime
},\epsilon ^{\prime })|J_{\alpha }^{\mathrm{tr}}|Z_{s}(p,\epsilon )\rangle }{%
(p^{2}-m^{2})(p^{\prime 2}-m_{X}^{2})}  \notag \\
&&\times \langle Z_{s}(p,\epsilon )|J_{\mu }^{^{\dagger }}|0\rangle +\ldots ,
\label{eq:Phys1}
\end{eqnarray}%
where contributions coming from the excited and continuum states are shown
by dots.

The physical side of the required sum rules can be expressed in terms of the
$Z_{s}$ and $X(4274)$ states' parameters, as well as matrix element $\langle
X(p^{\prime },\epsilon ^{\prime })|J_{\alpha }^{\mathrm{tr}%
}|Z_{s}(p,\epsilon )\rangle $ written down using weak transition form
factors $G_{i}(q^{2})$. The matrix elements of the $Z_{s}$ and $X(4274)$
states are rather simple:
\begin{equation}
\langle 0|J_{\nu }^{X}|X(p^{\prime },\epsilon ^{\prime })\rangle
=f_{X}m_{X}\epsilon _{\nu }^{\prime },  \label{eq:ME1}
\end{equation}%
and
\begin{equation}
\langle 0|J_{\mu }|Z_{s}(p,\epsilon )\rangle =fm\epsilon _{\mu }.
\label{eq:MElem2}
\end{equation}%
In Eqs.\ (\ref{eq:ME1}) and (\ref{eq:MElem2}) $f$ and $f_{X}$ are the
couplings of the states $Z_{s}$ and $X(4274)$, respectively. The vertex $%
\langle X(p^{\prime },\epsilon ^{\prime })|J_{\alpha }^{\mathrm{tr}%
}|Z_{s}(p,\epsilon )\rangle $ has more complicated expansion (see, Eq.\ (\ref%
{eq:Vertex})), and is modeled by means of the four universal transition form
factors $G_{i}(q^{2})$ which can be used for calculating all of the three
semileptonic decays.

Substituting the relevant matrix elements into Eq.\ (\ref{eq:Phys1}) we get
the final expression for $\Pi _{\mu \alpha \nu }^{\mathrm{Phys}}(p,p^{\prime
},q^{2})$
\begin{eqnarray}
&&\Pi _{\mu \alpha \nu }^{\mathrm{Phys}}(p,p^{\prime },q^{2})=\frac{%
fmf_{X}m_{X}}{(p^{2}-m^{2})(p^{\prime 2}-m_{X}^{2})}\left\{
G_{1}(q^{2})p_{\alpha }g_{\mu \nu }\right.   \notag \\
&&+G_{2}(q^{2})\left[ 1-\frac{m^{2}-m_{X}^{2}+q^{2}}{2m^{2}}\right] p_{\mu
}g_{\alpha \nu }  \notag \\
&&\left. -\frac{G_{3}(q^{2})}{2m^{2}}p_{\alpha }p_{\nu }p_{\mu }^{\prime
}+G_{4}(q^{2})\varepsilon _{\theta \alpha \mu \nu }p_{\theta }\right\}
+\ldots .  \label{eq:Phys2}
\end{eqnarray}%
By dots in $\Pi _{\mu \alpha \nu }^{\mathrm{Phys}}(p,p^{\prime },q^{2})$ we
denote not only effects due to the excited and continuum states, but also
contributions of structures which will not be used to derive the sum rules.

The QCD side of the sum rules can be found by employing $\Pi _{\mu \alpha
\nu }(p,p^{\prime })$ given by Eq.\ (\ref{eq:CF}), using the interpolating
currents and by contracting corresponding quark fields. These calculations
lead to $\Pi _{\mu \alpha \nu }^{\mathrm{OPE}}(p,p^{\prime },q^{2})$, $\ $%
expression of which in terms of the heavy and light $s-$quark propagators is
presented in the Appendix. In computations we use the $s$-quark and heavy
quark propagators given by the formulas
\begin{eqnarray}
&&S_{s}^{ab}(x)=i\delta _{ab}\frac{\slashed x}{2\pi ^{2}x^{4}}-\delta _{ab}%
\frac{m_{s}}{4\pi ^{2}x^{2}}-\delta _{ab}\frac{\langle \overline{s}s\rangle
}{12}  \notag \\
&&+i\delta _{ab}\frac{\slashed xm_{s}\langle \overline{s}s\rangle }{48}%
-\delta _{ab}\frac{x^{2}}{192}\langle \overline{s}g_{s}\sigma Gs\rangle
+i\delta _{ab}\frac{x^{2}\slashed xm_{s}}{1152}  \notag \\
&&\times \langle \overline{s}g_{s}\sigma Gs\rangle -i\frac{%
g_{s}G_{ab}^{\alpha \beta }}{32\pi ^{2}x^{2}}\left[ \slashed x{\sigma
_{\alpha \beta }+\sigma _{\alpha \beta }}\slashed x\right] +\ldots
\label{eq:qprop}
\end{eqnarray}%
and ($Q=b$ or $c$)
\begin{eqnarray}
&&S_{Q}^{ab}(x)=i\int \frac{d^{4}k}{(2\pi )^{4}}e^{-ikx}\Bigg \{\frac{\delta
_{ab}\left( {\slashed k}+m_{Q}\right) }{k^{2}-m_{Q}^{2}}  \notag \\
&&-\frac{g_{s}G_{ab}^{\alpha \beta }}{4}\frac{\sigma _{\alpha \beta }\left( {%
\slashed k}+m_{Q}\right) +\left( {\slashed k}+m_{Q}\right) \sigma _{\alpha
\beta }}{(k^{2}-m_{Q}^{2})^{2}}  \notag \\
&&\left. +\frac{g_{s}^{2}G^{2}}{12}\delta _{ab}m_{Q}\frac{k^{2}+m_{Q}{%
\slashed k}}{(k^{2}-m_{Q}^{2})^{4}}+\ldots \right\} ,  \label{eq:Qprop}
\end{eqnarray}%
and take into account terms up to dimension five.

The correlation function $\Pi _{\mu \alpha \nu }^{\mathrm{OPE}}(p,p^{\prime
},q^{2})$ contains the same Lorentz structures as its counterpart $\Pi _{\mu
\alpha \nu }^{\mathrm{Phys}}(p,p^{\prime },q^{2})$. We use the same
structures and corresponding invariant amplitudes to obtain the required sum
rules for the form factors $G_{i}(q^{2})$. But before that we make double
Borel transformation over variables $p^{2}$ and $p^{\prime 2}$ to suppress
contributions of the higher excited and continuum states, and perform
continuum subtraction. These rather routine manipulations give the sum rules
for the form factors $G_{i}(q^{2})$. For $G_{i}(q^{2}),$ $i=1\ $and $4$ we
get the similar sum rules%
\begin{eqnarray}
&&G_{i}(M^{2},\ s_{0},~q^{2})=\frac{1}{fmf_{X}m_{X}}\int_{\mathcal{M}%
_{1}^{2}}^{s_{0}}ds\int_{\mathcal{M}_{2}^{2}}^{s_{0}^{\prime }}ds^{\prime }
\notag \\
&&\times \rho _{i}(s,s^{\prime
},q^{2})e^{(m^{2}-s)/M_{1}^{2}}e^{(m_{X}^{2}-s^{\prime })/M_{2}^{2}},
\label{eq:FF1}
\end{eqnarray}%
where $M_{1}^{2},\ M_{2}^{2}$ are the Borel parameters, and $s_{0},\
s_{0}^{\prime }$ are the continuum threshold parameters that separate the
main contribution to the sum rules from the continuum effects. The limits of
the integrals in Eq.\ (\ref{eq:FF1}) and in expressions presented below are
defined in the form
\begin{equation}
\mathcal{M}_{1}^{2}=(m_{b}+m_{c}+2m_{s})^{2},\ \mathcal{M}%
_{2}^{2}=(2m_{c}+2m_{s})^{2}.
\end{equation}%
The remaining two sum rules read:%
\begin{eqnarray}
&&G_{2}(M^{2},\ s_{0},~q^{2})=\frac{2m}{f_{X}m_{X}f(m^{2}+m_{X}^{2}-q^{2})}
\notag \\
&&\times \int_{\mathcal{M}_{1}^{2}}^{s_{0}}ds\int_{\mathcal{M}%
_{2}^{2}}^{s_{0}^{\prime }}ds^{\prime }\rho _{2}(s,s^{\prime
},q^{2})e^{(m^{2}-s)/M_{1}^{2}}e^{(m_{X}^{2}-s^{\prime })/M_{2}^{2}},  \notag
\\
&&{}  \label{eq:FF2}
\end{eqnarray}%
and
\begin{eqnarray}
&&G_{3}(M^{2},\ s_{0},~q^{2})=-\frac{2m}{f_{X}m_{X}f}\int_{\mathcal{M}%
_{1}^{2}}^{s_{0}}ds\int_{\mathcal{M}_{2}^{2}}^{s_{0}^{\prime }}ds^{\prime }
\notag \\
&&\times \rho _{3}(s,s^{\prime
},q^{2})e^{(m^{2}-s)/M_{1}^{2}}e^{(m_{X}^{2}-s^{\prime })/M_{2}^{2}}.
\label{eq:FF3}
\end{eqnarray}%
As is seen the sum rules are written down using the spectral densities $\rho
_{i}(s,s^{\prime },q^{2})$ which are proportional to the imaginary part of
the corresponding invariant amplitudes in $\Pi _{\mu \alpha \nu }^{\mathrm{%
OPE}}(p,p^{\prime },q^{2})$. All of them contain both the perturbative and
nonperturbative contributions and are calculated with dimension-5 accuracy.
Their explicit expressions are very cumbersome, therefore we refrain from
providing them here. Sum rules for $G_{i}(q^{2})$ will be used in the next
section to find corresponding fit functions $F_{i}(q^{2})$ and calculate the
width of the semileptonic decays.


\section{Width of the decay $Z_{s}\rightarrow X(4274)\overline{l}\protect\nu %
_{l}$ and numerical results}

\label{sec:Decays}
The differential decay rate of the process $Z_{s}\rightarrow X(4274)%
\overline{l}\nu _{l}$ can be calculated using well known formulas: it is
given by the expression
\begin{eqnarray}
&&\frac{d\Gamma }{dq^{2}}=\frac{G_{F}^{2}|V_{cb}|^{2}}{3\cdot 2^{9}\pi
^{3}m^{3}}\left( \frac{q^{2}-m_{l}^{2}}{q^{2}}\right) \lambda \left(
m^{2},m_{X}^{2},q^{2}\right)  \notag \\
&&\times \left[ \sum_{i=1}^{i=4}G_{i}^{2}(q^{2})\mathcal{A}%
_{i}(q^{2})+G_{1}(q^{2})G_{2}(q^{2})\mathcal{A}_{12}(q^{2})\right.  \notag \\
&&\left. +G_{1}(q^{2})G_{3}(q^{2})\mathcal{A}%
_{13}(q^{2})+G_{2}(q^{2})G_{3}(q^{2})\mathcal{A}_{23}(q^{2})\right.\bigg] ,
\notag \\
&&  \label{eq:DifW}
\end{eqnarray}%
where
\begin{eqnarray*}
\lambda \left( m^{2},m_{X}^{2},q^{2}\right) &=&\left[ m^{4}+m_{X}^{4}+q^{4}%
\right. \\
&&\left. -2(m^{2}m_{X}^{2}+m^{2}q^{2}+m_{X}^{2}q^{2})\right] ^{1/2}.
\end{eqnarray*}%
In these calculations we neglect the mass of the neutrino $\nu _{l}$. The
decay rate $d\Gamma /dq^{2}$ depends on the transition form factors $%
G_{i}(q^{2}),$ and on functions $\mathcal{A}_{i}(q^{2})$ and $\mathcal{A}%
_{ij}(q^{2})$ explicit expressions of which are collected in the Appendix.
Therefore, \ as the first step in this situation we fulfil the calculation
of the form factors from sum rules and fit them by simple formulas which
allow us to perform integration over the whole region of momentum transfer $%
q^{2}$ and evaluate $\Gamma $.

Technical sides of numerical calculations in the context of the sum rules
approach are well known. Indeed, the sum rules given by Eqs.\ (\ref{eq:FF1}%
), (\ref{eq:FF2}) and (\ref{eq:FF3}) through the spectral densities $\rho
_{i}(s,s^{\prime },q^{2})$ depend on the quark, gluon and mixing
condensates, numerical values of which should be specified. Apart from these
input parameters they contain also masses and couplings of the tetraquarks $%
Z_{s}$ and $X(4274),$ as well as masses of the $s,$ $c$ and $b-$quarks.
\begin{widetext}

\begin{figure}[h!]
\begin{center} \includegraphics[%
totalheight=6cm,width=8cm]{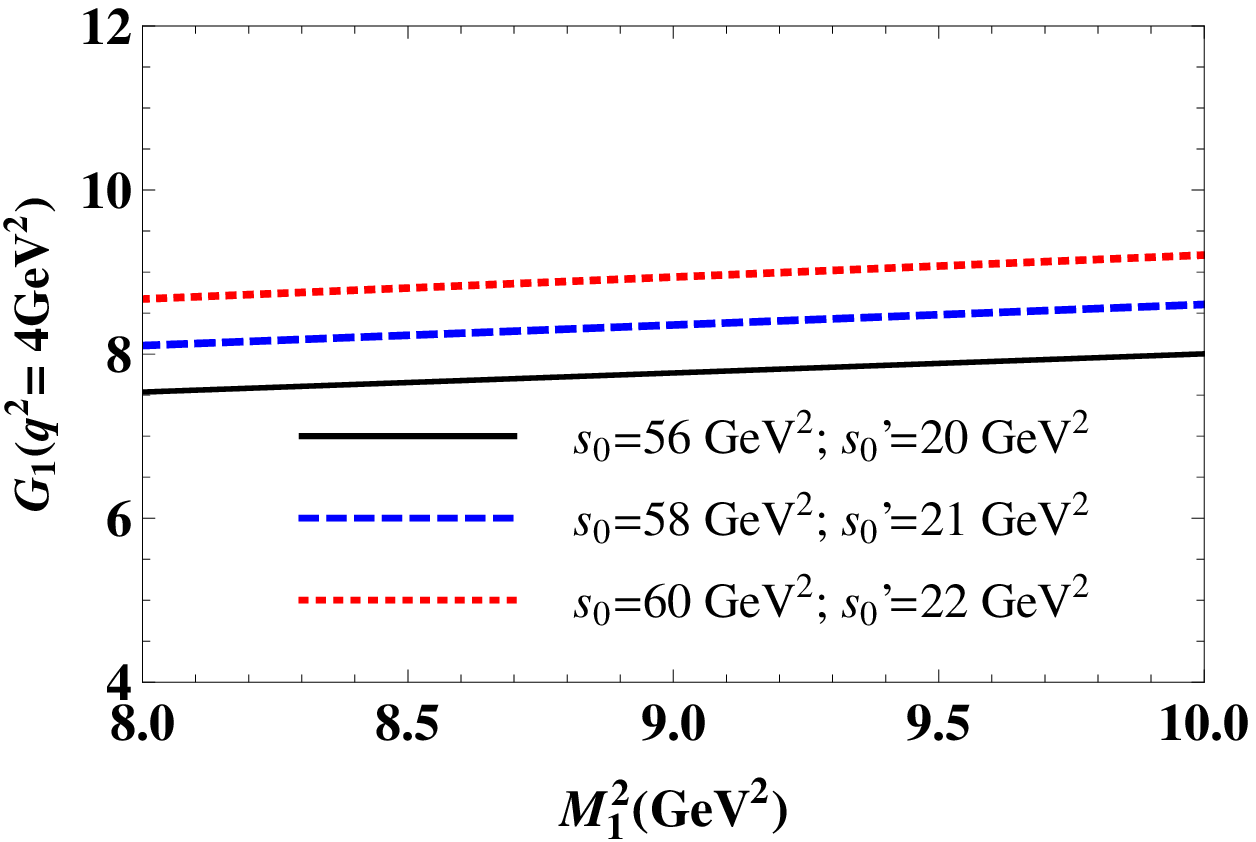}\,\,
\includegraphics[%
totalheight=6cm,width=8cm]{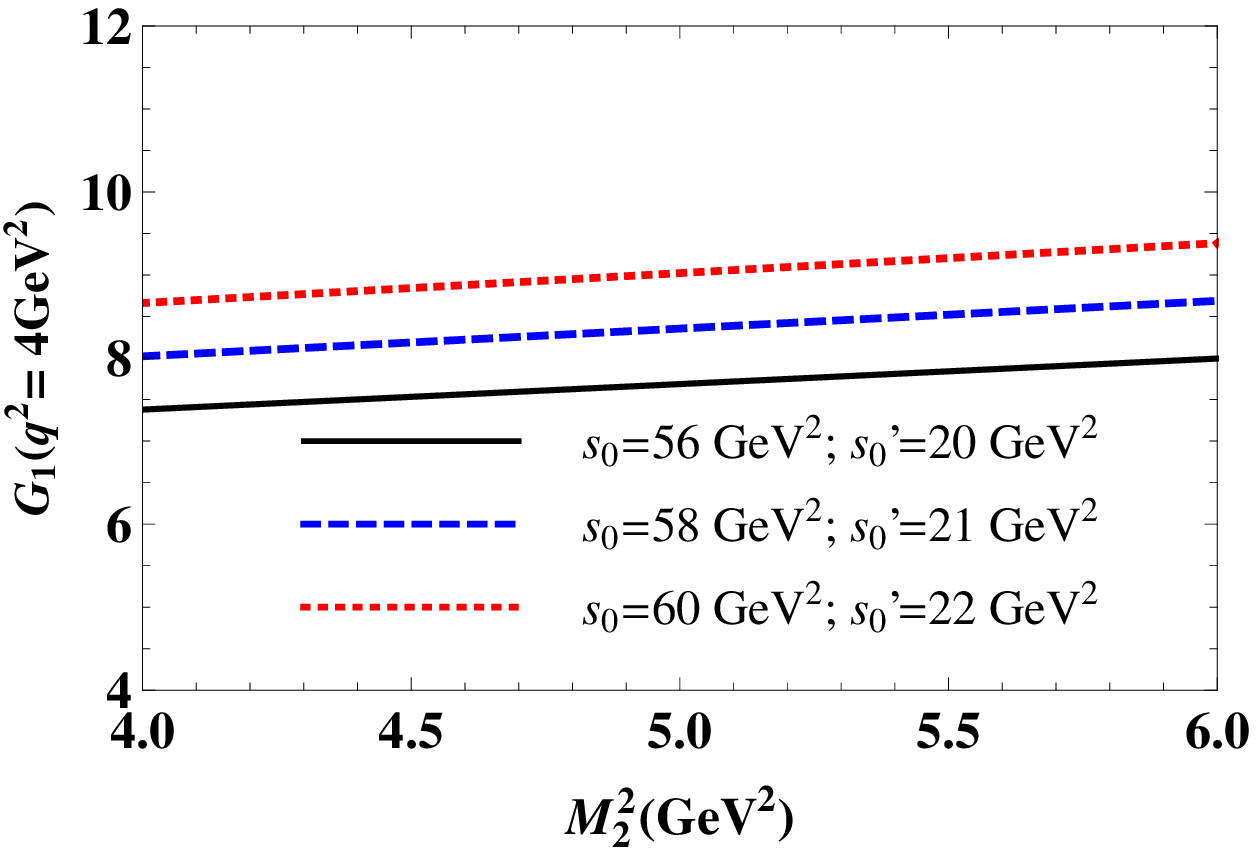}
\end{center}
\caption{ The form factor $G_1(q^2)$ at fixed $q^2=4\ \mathrm{GeV}$ as a function of
the Borel parameter $M_{1}^2$ (left panel), and as a function of the $%
M_{2}^2$ (right panel).}
\label{fig:G1}
\end{figure}

\end{widetext}

The spectroscopic parameters of the tetraquarks were evaluated in Refs.\
\cite{Agaev:2017uky} and \cite{Agaev:2017foq}: the mass of the $Z_{s}$ state
is given by Eq.\ (\ref{eq:Mass1}), and its coupling is equal to
\begin{equation}
f=(0.63\pm 0.19)\cdot 10^{-2}\ \mathrm{GeV}^{4}.  \label{eq:Coupl1}
\end{equation}%
The same parameters of the resonance $X(4274)$ read:%
\begin{eqnarray}
m_{X} &=&4264\pm 117~\mathrm{MeV},  \notag \\
f_{X} &=&(0.94\pm 0.16)\cdot 10^{-2}\ \mathrm{GeV}^{4}.  \label{CMass}
\end{eqnarray}%
The mass of the quarks are borrowed from Ref.\ \cite{Patrignani:2016xqp} $%
m_{s}=128\pm 10~\mathrm{MeV,\ }m_{c}=1.28\pm 0.03~\mathrm{GeV}$ and $%
m_{b}=4.18_{-0.03}^{+0.04}~\mathrm{GeV}$ (let us note that the mass of the $s
$-quark is rescaled to the normalization point $\mu _{0}^{2}=1\ \mathrm{GeV}%
^{2}$). For the Fermi coupling constant $G_{F}$ and CKM matrix element $%
|V_{bc}|$ we use:
\begin{eqnarray}
G_{F} &=&1.16637\cdot 10^{-5}\ \mathrm{GeV}^{-2},\   \notag \\
|V_{bc}| &=&(41.2\pm 1.01)\cdot 10^{-3}.  \label{eq:FermiA}
\end{eqnarray}%
Besides that we fix values of the quark, gluon and mixed local operators,
which contain important nonperturbative information. For these quantities we
utilize their well known values
\begin{eqnarray}
&&\langle \bar{q}q\rangle =-(0.24\pm 0.01)^{3}\ \mathrm{GeV}^{3},\ \langle
\bar{s}s\rangle =0.8\ \langle \bar{q}q\rangle ,  \notag \\
&&m_{0}^{2}=(0.8\pm 0.1)\ \mathrm{GeV}^{2},\ \langle \overline{q}g_{s}\sigma
Gq\rangle =m_{0}^{2}\langle \overline{q}q\rangle ,  \notag \\
&&\langle \overline{s}g_{s}\sigma Gs\rangle =m_{0}^{2}\langle \bar{s}%
s\rangle ,  \notag \\
&&\langle \frac{\alpha _{s}G^{2}}{\pi }\rangle =(0.012\pm 0.004)\,\mathrm{GeV%
}^{4}.  \label{eq:Param}
\end{eqnarray}%
Sum rules depend also on auxiliary parameters $M_{1}^{2}$, $M_{2}^{2}$ and $%
s_{0},\ s_{0}^{\prime }$ which should comply with standard constraints: at $%
M_{1,2,\mathrm{max}}^{2}$ prevalence of the pole contribution (\textrm{PC})
over other terms, and for $M_{1,2,\mathrm{min}}^{2}$ convergence of the
operator product expansion has to be satisfied. Minimal dependence of
evaluated quantities on the Borel parameters is also among the restrictions
that have has to be meet when choosing the domain $(s_{0},\ s_{0}^{\prime })$

For the initial tetraquark $Z_{s}$ channel we fix%
\begin{equation}
M_{1}^{2}\in \lbrack 8,\ 10]\ \mathrm{GeV}^{2},\ s_{0}\in \lbrack 56,\ 60]\
\mathrm{GeV}^{2},  \label{eq:Wind1}
\end{equation}%
which are very close to the intervals obtained in Ref.\ \cite{Agaev:2017uky}
from the analysis of the two-point sum rules. The same is true for the $%
M_{2}^{2}$ and $s_{0}^{\prime }$ which characterize in the process the final
tetraquark state $X(4274)$ (see, Ref.\ \cite{Agaev:2017foq})
\begin{equation}
M_{2}^{2}\in \lbrack 4,\ 6]\ \mathrm{GeV}^{2},\ s_{0}^{\prime }\in \lbrack
20,\ 22]\ \mathrm{GeV}^{2}.  \label{eq:Wind2}
\end{equation}%
In deriving the intervals in Eqs.\ (\ref{eq:Wind1}) and (\ref{eq:Wind2}) we
apply the following criteria: for the pole contribution
\begin{equation}
\mathrm{PC}=\frac{G_{i}(\ M_{\mathrm{max}}^{2},\ s,\ q^{2})}{G_{i}(\ M_{%
\mathrm{max}}^{2},\infty ,\ q^{2})}\geq 0.5,  \label{eq:Rest1}
\end{equation}%
for the contribution of $\mathrm{Dim}5$ term
\begin{equation}
\frac{G_{i}^{\mathrm{Dim5}}(M_{\mathrm{min}}^{2},\infty _{,}\ q^{2})}{%
G_{i}(M_{\mathrm{min}}^{2},\infty _{,}\ q^{2})}\leq 0.05.  \label{eq:Rest2}
\end{equation}%
Let us emphasize that we vary the parameters $M_{1}^{2}$ and $M_{2}^{2}$
independently, without any additional assumption about a functional relation
between them.

It is not difficult to see that in these domains of parameters the
constraints imposed on $G_{i}(M^{2},\ s_{0},~q^{2})$ are satisfied. In fact,
at maximal values of the Borel parameters the pole contribution to the sum
rule with $\rho _{1}(s,s^{\prime })$, for example, equals to $0.56$. At the
lower limits of the Borel parameters contribution of $\mathrm{Dim5}$ term
amounts to $1.5\%$ of the full result. The similar estimates are valid also
for the other sum rules, as well.

In Fig.\ \ref{fig:G1} we plot the form factor $G_{1}(q^{2})$ as a function
of the Borel parameters. As is seen, the predictions for $G_{1}(q^{2})$
contain a residual dependence both on the Borel and continuum threshold
parameters which is typical for sum rules calculations. Nevertheless, these
ambiguities that generate final errors remain within limits allowable for
such kind of computations.

In order to obtain the full width of the decay $Z_{s}\rightarrow X(4274)%
\overline{l}\nu _{l}$ one has to integrate the differential decay rate $%
d\Gamma /dq^{2}$ within allowed kinematical limits $m_{l}^{2}\leq q^{2}\leq
(m-m_{X})^{2}$. But in the case of $l=e,\ \mu $ leptons the lower limit of
the integral is considerably smaller than $1\ \mathrm{GeV}^{2}$, but the
perturbative calculations lead to reliable predictions for momentum
transfers $q^{2}>1\ \mathrm{GeV}^{2}$. Therefore, we use the usual recipe by
replacing the transition form factors in the whole integration region by fit
functions which for perturbatively allowed values of $q^{2}$ coincide with $%
G_{i}(q^{2})$.

\begin{figure}[h]
\includegraphics[width=7cm]{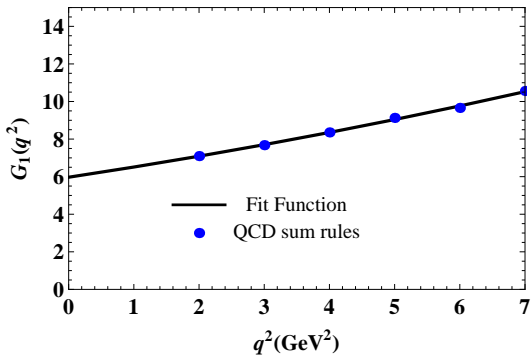}
\caption{The fit function $F_{1}(q^{2})$ for the transition form factor $%
G_{1}(q^{2})$.}
\label{fig:Fit}
\end{figure}

There are numerous analytical expressions for the fit functions. In the
present paper we use
\begin{equation}
F_{i}(q^{2})=f_{0}^{i}\exp \left[ c_{1i}\frac{q^{2}}{m_{\mathrm{fit}}^{2}}%
+c_{2i}\left( \frac{q^{2}}{m_{\mathrm{fit}}^{2}}\right) ^{2}\right] ,
\label{eq:FF}
\end{equation}%
where $f_{0}^{i},~c_{1i},\ c_{2i}$ and $m_{\mathrm{fit}}^{2}$ are fitting
parameters. In Fig.\ \ref{fig:Fit}, as an example, we depict the sum rules
results for the transition form factor $G_{1}(q^{2})$ and corresponding fit
function $F_{1}(q^{2})$. It is seen that Eq.\ (\ref{eq:FF}) leads to
reasonable agreement with QCD sum rules results. The fitting parameters for
all of the form factors are collected in Table \ref{tab:FitPar}.

\begin{table}[tbp]
\begin{tabular}{|c|c|c|c|c|}
\hline
$F_i(q^2)$ & $f_{0}^{i}$ & $c_{1i}$ & $c_{2i}$ & $m^2_{\mathrm{fit}}\ (%
\mathrm{GeV}^2)$ \\ \hline\hline
$F_{1}(q^2)$ & $5.97$ & $4.68$ & $-2.81$ & $53.29$ \\ \hline
$F_{2}(q^2)$ & $44.39$ & $1.39$ & $18.08$ & $53.29$ \\ \hline
$F_{3}(q^2)$ & $-10.50$ & $4.39$ & $1.78$ & $53.29$ \\ \hline
$F_{4}(q^2)$ & $-4.28$ & $7.01$ & $-20.73$ & $53.29$ \\ \hline
\end{tabular}%
\caption{The parameters of the fit functions used in evaluating  the $%
\Gamma $.}
\label{tab:FitPar}
\end{table}
As a result, for the full decay width of the processes $Z_{s}\rightarrow
X(4274)\overline{l}\nu _{l},\ l=e,\ \mu $ and $\tau $ we find%
\begin{eqnarray}
&&\Gamma \left( Z_{s}\rightarrow X\overline{e}\nu _{e}\right) =\left(
4.51\pm 1.56\right) \cdot 10^{-9\ \ }\mathrm{MeV},  \notag \\
&&\Gamma \left( Z_{s}\rightarrow X\overline{\mu }\nu _{\mu }\right) =\left(
4.47\pm 1.54\right) \cdot 10^{-9\ \ }\mathrm{MeV},  \notag \\
&&\Gamma \left( Z_{s}\rightarrow X\overline{\tau }\nu _{\tau }\right)
=\left( 9.03\pm 3.16\right) \cdot 10^{-10\ \ }\mathrm{MeV},  \label{eq:Width}
\end{eqnarray}%
which are the final results of the present investigations.

\section{Analysis and concluding notes}

\label{sec:CRel}
The width of the $Z_{s}$ tetraquark's dominant strong decay channel $\Gamma
(Z_{s}\rightarrow B_{c}\phi )=(168\pm 68)\ \mathrm{MeV}$ was evaluated in
Ref.\ \cite{Agaev:2017uky}. The $S$-wave decay $Z_{s}\rightarrow B_{c}\phi $
runs through the superallowed Okubo-Zweig-Iizuka mechanism, and constitutes
the main part of the $Z_{s}$ tetraquark's full width. Even neglecting its
other strong decays and comparing $\sim 168\ \mathrm{MeV}$ with widths from
Eq.\ (\ref{eq:Width}) one can see that semileptonic transitions of $Z_{s}$
are rare processes. Smallness of these decay widths is connected mainly with
the CKM matrix element $|V_{bc}|$. The width of the $Z_{s}$ tetraquark's
semileptonic decay that run through weak transition $c\rightarrow s+W^{+}$
owing to $|V_{cs}|$ may be larger approximately by a factor $10^{3}$ than $%
\Gamma \lbrack Z_{s}\rightarrow X(4274)\overline{l}\nu _{l}]$, but then the
final tetraquark is a state with unknown properties and parameters, which
should be explored separately. The process $Z_{s}\rightarrow X(4274)%
\overline{l}\nu _{l}$ can manifest itself through the decay chain $%
Z_{s}\rightarrow X(4274)\overline{l}\nu _{l}\rightarrow J/\psi \phi
\overline{l}\nu _{l}$ with two conventional mesons in the final state. It is
evident that the process $Z_{s}\rightarrow J/\psi \phi \overline{l}\nu _{l}$
is also a rare decay channel of the tetraquark $Z_{s}$.

Nevertheless, considered processes may provide valuable information about
the structure of the resonances $Z_{s}$, $X(4140)$ and $X(4274)$. In fact,
as we have emphasized above the states $X(4140)$ and $X(4274)$ have the same
spin-parities, and presumably are members of the color triplet and sextet
multiplets, respectively. Our investigations demonstrate that the open
charm-bottom color sextet tetraquark $Z_{s}$ can decay to the color sextet
resonance $X(4274)$ through the process $Z_{s}\rightarrow X(4274)\overline{l}%
\nu _{l}$, whereas the matrix element of the semileptonic transition $%
Z_{s}\rightarrow X(4140)\overline{l}\nu _{l}$ is identically equal to zero.

There may in general exist the open charm-bottom tetraquarks $Z_{s}^{\prime }
$ with triplet color structure $\left[ \overline{3}_{c}\right] _{sc}\otimes %
\left[ 3_{c}\right] _{\overline{s}\overline{b}}$, which constitute another
multiplet of open charm-bottom states. The spectroscopic parameters of these
tetraquarks should differ from those of the $Z_{s}$ states: this was proved
in the case of the resonances $X(4140)$ and $X(4274)$. It is not difficult
to demonstrate that an axial-vector state $Z_{s}^{\prime }=[cs][\overline{b}%
\overline{s}]$ with triplet color structure and interpolating current
\begin{equation}
J_{\mu }^{\prime }(x)=s_{a}^{T}C\gamma _{5}c_{b}\left( \overline{s}%
_{a}\gamma _{\mu }C\overline{b}_{b}^{T}-\overline{s}_{b}\gamma _{\mu }C%
\overline{b}_{a}^{T}\right)   \label{eq:Curr3}
\end{equation}%
decays to the final state $X(4140)\overline{l}\nu _{l}$. At the same time
the matrix element of transition $Z_{s}^{\prime }\rightarrow X(4274)%
\overline{l}\nu _{l}$ is identically equal to zero. In other words, the weak
interactions preserve the color structure of the involved axial-vector
tetraquarks: weak transitions from color triplet to sextet and from color
sextet to triplet states are forbidden. Hence, semileptonic decays
considered in the present work may clarify the underlying structure both of
the initial $Z_{s}$ and final tetraquarks. The multiplet of color triplet
open charm-bottom tetraquarks $Z_{s}^{\prime }$, their spectroscopic
parameters, strong and semileptonic decays deserve detailed investigations,
but this task is beyond the scope of the present work.

It is instructive also to compare mechanisms of tetraquarks' and
conventional mesons' hadronic and semileptonic decay modes. The hadronic
decays of tetraquarks to two conventional mesons are their dominant decay
channels. The reason is that the tetraquarks are resonances composed of
four-quarks and their strong transitions to two ordinary mesons do not
require a creation of additional quark-antiquark pair which is necessary in
decays of conventional mesons built of two valence quarks. Therefore, in the
lack of a gluon exchange these channels do not suffer from the corresponding
suppression. On the contrary, the semileptonic decays of these particles
proceed through the weak transition of a initial quark to a final quark and
the weak boson, and it is the same for both the tetraquarks and conventional
mesons: the difference between them is connected only with a number of the
spectator quarks. Therefore, experimental studies of semileptonic and strong
decays of resonances which are candidates to exotic states may give an
interesting information on nature of master particles: the relevant problems
deserve further detailed analysis.


\section*{ACKNOWLEDGEMENTS}

H. S. and B. B. thank T.~M.~Aliev for helpful discussions. H. S., B.~B. and
K.~A. appreciate financial support by TUBITAK through Grant No: 115F183.

\appendix*

\section{ The correlation function $\Pi _{\protect\mu \protect\alpha \protect%
\nu }^{\mathrm{OPE}}(p,p^{\prime },q^{2}),$ and the functions $\mathcal{A}%
_{i}\mathcal{(}q^{2}\mathcal{)},\ \mathcal{A}_{ij}\mathcal{(}q^{2}\mathcal{)}
$}

\renewcommand{\theequation}{\Alph{section}.\arabic{equation}} \label{sec:App}
In this Appendix we have collected the formulas for the correlation function
$\Pi _{\mu \alpha \nu }^{\mathrm{OPE}}(p,p^{\prime },q^{2})$ in terms of the
quark propagators, as well as explicit expressions of the functions $%
\mathcal{A}_{i}\mathcal{(}q^{2}\mathcal{)},\ \mathcal{A}_{ij}\mathcal{(}q^{2}%
\mathcal{)} $ that enter to expression of the differential decay rate $%
d\Gamma/dq^2$.
\begin{widetext}

\begin{eqnarray}
&&\Pi _{\mu \alpha \nu }^{\mathrm{OPE}}(p,p^{\prime },q^{2})=i^{2}\int
d^{4}xd^{4}ye^{-ipx}e^{ip^{\prime }y}\left\{ \mathrm{Tr}\left[ \gamma _{5}%
\widetilde{S}_{s}^{aa^{\prime }}(y-x)\gamma _{5}S_{c}^{bb^{\prime }}(y-x)%
\right] \left\{ \mathrm{Tr}\left[ \gamma _{\nu }\widetilde{S}%
_{c}^{ib}(-y)(1-\gamma _{5})\gamma _{\alpha }\widetilde{S}_{b}^{a^{\prime
}i}(x)\gamma _{\mu }S_{s}^{b^{\prime }a}(x-y)\right] \right. \right.   \notag
\\
&&+\mathrm{Tr}\left[ \gamma _{\nu }\widetilde{S}_{c}^{ib}(-y)(1-\gamma
_{5})\gamma _{\alpha }\widetilde{S}_{b}^{b^{\prime }i}(x)\gamma _{\mu
}S_{s}^{a^{\prime }a}(x-y)\right] +\mathrm{Tr}\left[ \gamma _{\nu }%
\widetilde{S}_{c}^{ia}(-y)(1-\gamma _{5})\gamma _{\alpha }\widetilde{S}%
_{b}^{a^{\prime }i}(x)\gamma _{\mu }S_{s}^{b^{\prime }b}(x-y)\right]   \notag
\\
&&\left. +\mathrm{Tr}\left[ \gamma _{\nu }\widetilde{S}_{c}^{ia}(-y)(1-%
\gamma _{5})\gamma _{\alpha }\widetilde{S}_{b}^{b^{\prime }i}(x)\gamma _{\mu
}S_{s}^{a^{\prime }b}(x-y)\right] \right\} +\mathrm{Tr}\left[ \gamma _{5}%
\widetilde{S}_{s}^{aa^{\prime }}(y-x)\gamma _{\nu }S_{c}^{bb^{\prime }}(y-x)%
\right] \left\{ \mathrm{Tr}\left[ \gamma _{5}\widetilde{S}%
_{c}^{ib}(-y)(1-\gamma _{5})\gamma _{\alpha }\widetilde{S}_{b}^{a^{\prime
}i}(x)\right. \right.   \notag \\
&&\left. \times \gamma _{\mu }S_{s}^{b^{\prime }a}(x-y)\right] +\mathrm{Tr}%
\left[ \gamma _{5}\widetilde{S}_{c}^{ib}(-y)(1-\gamma _{5})\gamma _{\alpha }%
\widetilde{S}_{b}^{b^{\prime }i}(x)\gamma _{\mu }S_{s}^{a^{\prime }a}(x-y)%
\right] +\mathrm{Tr}\left[ \gamma _{5}\widetilde{S}_{c}^{ia}(-y)(1-\gamma
_{5})\gamma _{\alpha }\widetilde{S}_{b}^{a^{\prime }i}(x)\gamma _{\mu
}S_{s}^{b^{\prime }b}(x-y)\right]   \notag \\
&&\left. \left. +\mathrm{Tr}\left[ \gamma _{5}\widetilde{S}%
_{c}^{ia}(-y)(1-\gamma _{5})\gamma _{\alpha }\widetilde{S}_{b}^{b^{\prime
}i}(x)\gamma _{\mu }S_{s}^{a^{\prime }b}(x-y)\right] \right\} \right\} ,
\label{eq:A1}
\end{eqnarray}

\end{widetext}
where
\begin{equation*}
\widetilde{S}_{s(b,c)}(x)=CS_{s(b,c)}^{T}(x)C,
\end{equation*}%
Here $S_{s(b,c)}(x)$ are $s,$ $b$ and $c$ quarks' propagators, explicit
formulas of which have been written down in the main text of the paper.

The functions $\mathcal{A}_{i}(q^{2})$ and $\mathcal{A}_{ij}(q^{2})$ are
determined by the expressions:
\begin{widetext}

\begin{equation*}
\mathcal{A}_{1}(q^{2})=\frac{1}{m^{2}m_{X}^{2}q^{4}}\left[ m_{X}^{4}+%
\widetilde{m}_{1}^{4}+2m_{X}^{2}\left( 4m^{2}+\widetilde{m}_{1}^{2}\right) %
\right] \left[ q^{4}m_{l}^{2}\left( 2m_{X}^{2}+m^{2}+\widetilde{m}%
_{1}^{2}\right) -m_{l}^{4}\widetilde{m}_{1}^{4}+q^{4}\left( m_{X}^{4}+%
\widetilde{m}_{1}^{4}-2m_{X}^{2}m_{1}^{2}\right) \right] ,
\end{equation*}%
\begin{equation*}
\mathcal{A}_{2}(q^{2})=\frac{(q^{2}-m_{l}^{2})}{m^{2}m_{X}^{2}q^{2}}\left(
m_{X}^{4}+\widetilde{m}_{1}^{4}-2m_{X}^{2}m_{1}^{2}\right) \left[
m_{l}^{2}m_{2}^{2}+q^{2}(m_{2}^{2}+q^{2})\right] ,
\end{equation*}%
\begin{equation*}
\mathcal{A}_{3}(q^{2})=\frac{1}{16m^{6}m_{X}^{2}q^{4}}\left( m_{X}^{4}+%
\widetilde{m}_{1}^{4}-2m_{X}^{2}m_{1}^{2}\right) ^{2}\left[
q^{4}m_{l}^{2}(2m_{2}^{2}-q^{2})-m_{l}^{4}\widetilde{m}_{2}^{4}+q^{4}\left(
m_{X}^{4}+\widetilde{m}_{1}^{4}-2m_{X}^{2}m_{1}^{2}\right) \right] ,
\end{equation*}%
\begin{eqnarray*}
\mathcal{A}_{4}(q^{2}) &=&\frac{1}{m^{2}m_{X}^{2}q^{4}}\left[ 8m^{2}m_{X}^{2}%
\widetilde{m}%
_{2}^{4}q^{4}+m_{2}^{2}(m_{2}^{4}+12m^{2}m_{X}^{2})q^{6}-2(m_{2}^{4}+4m^{2}m_{X}^{2})q^{8}+m_{2}^{2}q^{10}+8m_{l}^{2}m^{2}m_{X}^{2}(q^{2}-2m_{2}^{2})q^{4}\right.
\\
&&\left. -m_{l}^{4}(8m^{2}m_{X}^{2}\widetilde{m}_{2}^{4}+q^{2}m_{2}^{2}%
\widetilde{m}_{2}^{4}+2q^{4}m_{2}^{4}-q^{6}m_{2}^{2})\right] ,
\end{eqnarray*}%
\begin{equation*}
\mathcal{A}_{12}(q^{2})=\frac{2(m_{l}^{2}-q^{2})}{m^{2}m_{X}^{2}}\left[
m_{X}^{4}(m_{1}^{2}+2q^{2})-m_{X}^{6}-\widetilde{m}_{1}^{6}+m_{X}^{2}\left(
m^{4}+2q^{2}m^{2}-3q^{4}\right) \right] ,
\end{equation*}%
\begin{eqnarray*}
\mathcal{A}_{13}(q^{2}) &=&\frac{1}{2m^{4}m_{X}^{2}q^{4}}\left[ m_{X}^{6}+%
\widetilde{m}_{1}^{6}-m_{X}^{4}(m_{1}^{2}+2q^{2})-m_{X}^{2}\left(
m^{4}+2m^{2}q^{2}-3q^{4}\right) \right] \left[
m_{l}^{2}q^{4}(2m_{1}^{2}-q^{2})-m_{l}^{4}\widetilde{m}_{2}^{4}\right.  \\
&&\left. +q^{4}\left( m_{X}^{4}+\widetilde{m}_{1}^{4}-2m_{X}^{2}m_{1}^{2}%
\right) \right] ,
\end{eqnarray*}%
\begin{equation*}
\mathcal{A}_{23}(q^{2})=\frac{q^{2}-m_{l}^{2}}{2m^{4}m_{X}^{2}}\left(
m_{X}^{4}+\widetilde{m}_{1}^{4}-2m_{X}^{2}m_{1}^{2}\right) ,
\end{equation*}%
where%
\begin{equation*}
m_{1}^{2}=m^{2}+q^{2},\widetilde{m}_{1}^{2}=m^{2}-q^{2},\
m_{2}^{2}=m^{2}+m_{X}^{2},\ \widetilde{m}_{2}^{2}=m^{2}-m_{X}^{2}\ .
\end{equation*}

\end{widetext}

\end{document}